\documentclass[3p,review]{elsarticle}

\usepackage{lineno,hyperref,lscape,url}

\begin{document}

\begin{frontmatter}

\title{Modeling of raining season onset and cessation of tropical rainfall for climate change adaptation in Agriculture}

\author[futa,lokoja]{Ibiyinka A. Fuwape}
\ead{iafuwape@futa.edu.ng}

\author[futa]{Samuel T. Ogunjo\corref{cor1}}
\ead{stogunjo@futa.edu.ng}
\cortext[cor1]{Corresponding author}

\address[futa]{Department of Physics, Federal University of Technology, Akure}
\address[lokoja]{Department of Physics, Federal University, Lokoja}

\begin{abstract}

This study investigates the trend in Rainfall Onset Dates (ROD), Rainfall Cessation Dates (RCD), Length of Growing Seasons (LGS) and Rainfall Amount at Onset of Rainfall (RAO) using linear regression, Mann-Kendall, Sen Slope and Hurst Exponent for four locations in tropical Nigeria and the development of a Fourier based model for ROD and RCD.  Daily data was obtained from the Nigerian Meteorological Agency for thirty-four (34) years (1979 - 2013).  ROD and RCD were computed using the method of cumulative percentage mean rainfall values.  Maiduguri, Gusau and Ikom showed positive trends in ROD and RCD while Ibadan exhibited negative trends in the two parameters.  Anti-persistence was observed in ROD, RCD and LGS for three locations (Maiduguri, Gusau and Ibadan).  A Fourier based model with seven (7) coefficients was developed to model ROD and RCD for all the locations.  The model developed performed very well in all locations with the best performance obtained in Gusau and Ibadan for ROD and RCD respectively.  The effects of climate change on agricultural output for the four (4) locations under consideration were highlighted and adaption techniques suggested for mitigating the impact on agricultural output and livelihood of citizens in the areas.
\end{abstract}


\end{frontmatter}

\newpage
\section{Introduction}

Climate play an important role in different aspect of national economy such as health, agriculture, tourism, and others, directly or indirectly.  The role of weather and climate has been reported in health \citep{Adeniji-Oloukoi2013,Fuhrmann2010,Ermert2012}, power generation \citep{parkpoom,hor} , tourism \citep{martin}, tropospheric communication \citep{Fuwape2016} and other sectors of the economy.  However, the most directly affected sector of the economy is agriculture.  Agriculture and agricultural practices in sub-saharan Africa depends on elements of climate such as rainfall and sunshine due to its location.  Nigeria, like many other sub-saharan countries, has an agrarian economy.  The agricultural sub-sector contributed about $20.24 - 48.57	\%$ to the Nigerian economy between 1981 and 2015.  (\url{http://data.worldbank.org/indicator/NV.AGR.TOTL.ZS}).  In the first quarter of 2016, Agriculture contributes 23\% to the national Gross Domestic Product (GDP) and employs 70\% of the workforce with 85\% involves in crop production.   A significant change in climate will greatly affect agricultural output, sources of livelihood of many citizens and the Gross Domestic Product (GDP) of the country. Hence, weather and climate are important factors to consider in sub-saharan Africa whose economy depends heavily on agriculture and agriculture related outputs.  It is important to study how the climate is changing over time to better prepare and mitigate its effect on our economy.   As part of the Sustainable Development Goals,  the United Nations is to ``take urgent action to combat climate change and its impact" by taking steps such as "integrate climate change measures into national policies, strategies and planning".

Current adaptation methods by farmers such as changing planting times, improved farming practices or better crop species will not suffice if the rate of change is faster than speculated.  There is currently no coordinated plan for mitigation of climate change effect by both the Federal and State Governments of Nigeria, except Lagos State.   The response of Lagos State, Southwestern Nigeria to climate change includes: establishment of special purpose organizations such as Lagos State Waste Management Authority (LAWMA) and Lagos State Emergency Management Agency (LASEMA),  improved sea wall protection, amongst others \citep{Elias2015}.  However, none of the stated and implemented initiatives will adequately address agricultural output in the State, as the environment is the focus of the current policy.  A report by  \citet{Burke2009} indicated that several countries will have novel future climates similar to the climate of at least five other countries.  The expected future climate of Nigeria, according to the report, is posited to accommodate crops that are currently poorly represented in major genebanks. Thus, detailed study of how our climate is changing, what it is likely to change to and methods to adopt for seamless agricultural adaptation is essential at this time.


Due to different sources and length of data, computational methods and number of locations used, differing results have been reported by different researchers for rainfall trend over Nigeria.   \citet{Adefolalu1986} observed that there is a decline in the amount of monthly rainfall which is more significant in the north with the central parts of the country exposed more to drought than other parts of the country.  Out of twenty (20) locations investigated for rainfall trend in Nigeria, fifteen (15) showed increasing trend and five (5) showed decreasing trend while cycles of 1-2, 7-10 and 15 years were reported \citep{Alli2012}.  \citet{Oguntunde2011}, using rainfall variability index, Mann-Kendall trend test, reported that 90\% of the entire country show negative trend in rainfall amount with 22\% of the locations showing significant trends.  Dominant peaks with periods 2-3, 5-7, 10-15 and 30 yrs were reported and accounted for.  A recent study by \citet{Ogunjo2016}, using deterministic chaos, suggests that rainfall is more predictable in the north than in the southern part of Nigeria.  For a more informative research that can help farmers in the country, it is important to study other variables that affect planting, crop selection, harvest and other aspect of farming in addition to seasonal and annual rainfall amount.

One of the major determinants of suitable crops for a region is the length of the growing season (LGS) which is determined by the rain onset dates (ROD) and rain cessation dates (RCD).  `Growing season` can be defined as the period of the year during which rainfall distribution characteristics are suitable for crop germination, establishment, and full development \citep{Odekunle2004}. Seeds require appropriate amount of rainfall and temperature for successful germination, hence, the timing of planting season is important for good planting season.  Different definitions of ROD have been proposed for the West African region based on rainfall amount, Inter Tropical Discontinuity (ITD)-Rainfall, rainfall-evaporation relationship, wind-shear scheme, Theta-E method \citep{Odekunle2005}.  The use of rainfall amount for computation of rainfall onset and retreat is the most popular with various techniques developed based on it.

Onset and cessation of rainfall has been studied in different parts of the country.  In a study by \citet{Oguntunde2014} using thirty-two (32) stations over Nigeria, negative trends were reported in 34.4\%, 31.2\%, 50.0\% for onset, cessation and length of growing seasons while positive trends were observed in 59.4\%, 40.6\%, 34.4\%  of the locations studied.  The dependence of rainfall on upper atmosphere wind parameters led \citet{Omotosho1992} to develop an empirical model for long term prediction of onset and cessation of rainfall in the Sahelian climate based on wind shear direction in the troposphere.   Using a simple water balance method, negative trends were reported for onset of rainfall in Northern Nigeria with no significant trends but statistically significant negative trend was observed for cessation and length of growing season for the same locations \citep{Oladipo1993}.  \citet{Thomson2010} developed nonparametric equations for the prediction of onset and cessation of rainfall using the following climatic indices: El Nino-Southern Oscillation, East Pacific Oscillation, Northern Annular Mode, North Atlantic Oscillation, Northern Oscillation Index, Tropical Northern Atlantic Index, Tropical Southern Atlantic Index, Pacific Decadal Oscillation, Pacific/North American Pattern and West Pacific Oscillations for selected locations in Nigeria.   \citet{Omotosho2000} proposed a long range empirical model based on potential temperature for onset and cessation of rainfall in selected locations in West Africa. A multivariate model for onset and cessation rainfall for selected locations in Nigeria using sea surface temperature (SST), land surface temperature (LST) and location of the Inter Tropical Discontinuity (ITD) was developed by \citet{Odekunle2005}.  \citet{Laux2008} developed a fuzzy method for detection of onset of rainfall and a method of predicting onset based on onset days for other regions in West Africa.  \citet{Adejuwon1990} reported a decreasing trend in all of the stations and most of the stations for cessation and onset of rainfall respectively.  Most of these models developed for the computation of rainfall onset and cessation are based on atmospheric parameters which are not readily available in the region (potential temperature, wind shear temperature).

Previous researches on onset and cessation of rainfall have employed linear trends.  In this study, spectral peaks in onset and cessation of rainfall are investigated.  The observed peaks are used to estimate periodicity in the parameters, hence, a model that will account for the linear and periodic trend is developed for the study areas.
The specific objectives of this current research are to: identify, if any, trend (linear and periodic) in onset and cessation of rainfall using linear and nonlinear methods, develop a model for the periodic and linear trend identified, and propose ways of mitigating the effect of climate change on agricultural sector of the regions studied

\section{Study Area and Data}
Nigeria is a tropical country located between latitude $4 - 14^oN$ and longitude $2-15^oE$.  The climate varies widely between the southern part which is bordered by the Atlantic Ocean and the Northern part bordered by Chad and Niger Republic.  The varying climate results in diverse vegetation and agricultural practices within the country.  The climate is governed by the tropical continental (cT) air mass from the Sahara desert and the tropical air mass from the South Atlantic Ocean.  These two meet at a boundary referred to as the Inter-Tropical discontinuity (ITD).   The country has two distinct seasons: dry season from November to March and a wet season from April to October.  The country experiences a phenomenon referred to as the `little dry season' in the month of August.  Nigeria has a large percentage of arable land with annual rainfall of 830 - 1450 mm/year \citep{Oguntunde2011}.

\begin{figure}
  \centering
  \includegraphics[scale=0.6]{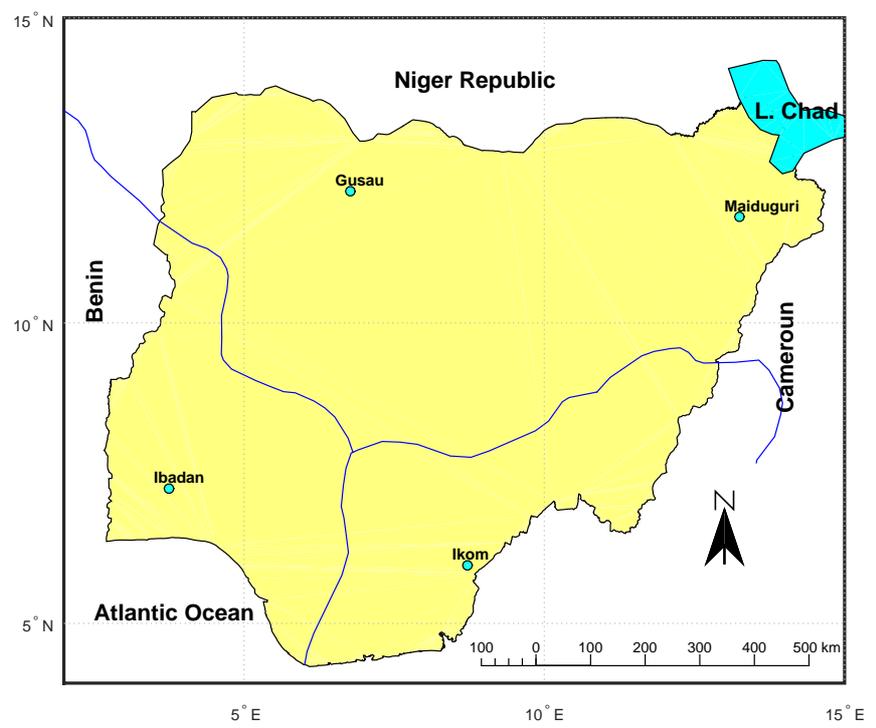}\\ 
  \caption{Map of Nigeria showing the study areas}\label{fig1}
\end{figure}

Two locations  (Gusau and Maiduguri) from the northern part and two locations (Ibadan and Ikom) from the southern region were chosen for this study (Figure \ref{fig1}) to capture the extreme climates in the country.  Ikom is located in Cross River State, South Eastern part of the country.  The main food crops grown in the State are tubers such as Yams, Cassava and cocoyam.    Maiduguri is the capital of Borno State, North Eastern Nigeria.  The location is suitable for grains such as Millet, Sorghum, Groundnut, Beans, Maize and Rice. Maiduguri is also known for the animal husbandry practiced there with domestic animals such as cows, sheep and goats.  Recently, agricultural activites in Borno State is being restored after some years of terrorist related incidents.  Gusau (Zamfara State) located in the North Western region of the country share the same agricultural activities with Maiduguri.  Ibadan, located in Oyo State, South Western Nigeria was also chosen for this study.  In Ibadan, more of tubers (Yam and Cassava) are produced than grains (Sorghum and rice).   The location also has the highest number of poultry farms in the country.

Daily rainfall data for this study was obtained from the archives of the Nigerian Meteorological Agency (NIMET), Lagos, Nigeria for thirty-four (34) years (1979 - 2013).  Quality assurance of the data was carried out by the staff of the agency in line with best practices.
\section{Data Analysis}
\subsection{Computation of ROD,   RCD, 	RAO and	 LGS}
The onset and cessation of rainfall was computed by a rainfall based method proposed by \citet{Odekunle2005}.  The method is a direct method as it uses rainfall and uses rainfall data which is readily available in different regions of the country.  \citet{Odekunle2004} defined the onset of rainfall as the period of the year in which $7-8\%$ mean cumulative rainfall over a 5-day period is attained and the cessation of rainfall as the period of the year when 90\% of the mean cumulative rainfall is attained.  The mean annual rainfall over a 5-day period ($y_i = \sum_{i}^{i+4} r_i$), where $i=1,2,3,\cdots, n-5$, n is the number of days in the year and r is the rainfall amount. The percentage occurrence of $y_i$ in a year is then computed

\begin{equation}\label{onset}
    f_j = \frac{y_i}{\sum y_i} \times 100
\end{equation}

The cumulative sum of $x_k$ is then calculated using the expression
\begin{equation}\label{cumsum}
    x_k = \sum_{k=1}^{n-1} f_k + f_{k+1}
\end{equation}

The point that correspond to $7-8\%$ and $90\%$ on the graph  $x_k$ against time for each year are the ROD and RCD respectively \citep{Odekunle2006}.  In this work, LGS is computed as the difference between ROD and RCD while RAO is defined as the amount of rainfall available at ROD.
\subsection{Trend Analysis}
The linear trend was computed using simple linear regression of the form $y = ax + b$, where x and y are the independent and dependent variable respectively, a and b are constants to be determined by minimizing the function
\begin{equation}\label{ltrend}
Q = \sum_{i=1}^n \left( y_i - (ax_i + b) \right) = 0
\end{equation}
The slope of the straight line is an indication of the trend in the time series.

Mann-Kendall test is a popular test for trends in meteorological data sets as it does not assume normality of the time series.  The Mann-Kendall test gives the direction but not the magnitude of the significant trends. The Mann-Kendall test statistics S is based on the pair-wise comparison of each data points with all preceding data points.

\begin{equation}\label{mkend}
    S = \sum_{k=1}^{n-1}\sum_{j=k+1}^n sgn (x_j - x_k)
\end{equation}
where $n$ is the length of the time series $x_1,\cdots,x_n$, $sgn(\cdot)$ is a sign function while $x_j$ and $x_k$ are values in years.   The variance of S is computed as:
\begin{equation}\label{varKen}
    \sigma^2(S) = \frac{1}{18} \left[n(n-1)(2n+5) - \sum_{p=1}^q t_p(t_p-1)(2t_p + 5)  \right]
\end{equation}
where $q$ is the number of tied groups and $t_p$ is the number of data values in the $p^{th}$ group \citep{Oguntunde2011,Olofintoye2010}.  The test statistic Z obtained as:

\begin{equation}\label{ZKend}
   Z = \left\{
      \begin{array}{ll}
        \frac{S-1}{\sqrt{\sigma^2(S)}}, & \hbox{if $S > 0$;} \\
           & \\
        0, & \hbox{if $S = 0$;} \\
           & \\
        \frac{S+1}{\sqrt{\sigma^2(S)}}, & \hbox{if $S < 0$.}
      \end{array}
    \right.
\end{equation}

If the trend of a time series is assumed to be linear and of the form $f(t) = Qt + B$, Sen slope (Q) can be estimated from the expression
\begin{equation}\label{senss}
    Q_i = \frac{x_j - x_k}{j-k}
\end{equation}
where $j>k$.  The Sen estimate of the slope is the median value of $Q_i$ \citep{Olofintoye2010}.
\subsection{Hurst Exponent}
The Hurst exponent can be
used to determine whether a time series is random or not. The exponent shows persistence, anti-persistence \citep{Fuwape2015}
and predictability \citep{ogunjo2015effect} in a time series.  From Equation \ref{hurst1}, Hurst Exponent values between 0 and 0.5 shows anti-persistence which implies that increasing trend will likely be followed by decreasing trend and vice versa.  Values of Hurst Exponent between 0.5 and 1, shows persistence, that is,  increasing trend will be followed by increasing trend and vice versa \citep{Fuwape2015}.  Many algorithms have been proposed for the computation of Hurst Exponent, however in this work, the method of Detrended Fluctuation Analysis will be employed.
\begin{equation}\label{hurst1}
   H = \left\{
      \begin{array}{ll}
        0 < H < 0.5, & \hbox{anti-persistence;} \\
           & \\
        0.5, & \hbox{random walk;} \\
           & \\
        0.5 < H < 1, & \hbox{persistence.}
      \end{array}
    \right.
\end{equation}
The method of Detrended Fluctuation Analysis was introduced by \citet{Peng1994}.  In this approach, the fluctuation $F(t)$ of N segements of length $\tau$ of a time series is computed
\begin{equation}\label{hurst2}
    F(\tau) = \frac{1}{\tau}\sum_{t=1}^\tau (y(t) - Y_m(t))^2
\end{equation}
where, $Y_m(t)$ is a local m-polynomial trend within the specified segment \citep{ogunjo2015effect}.   The Hurst exponent, H, is obtained from the expression
\begin{equation}\label{hurst3}
    F(\tau) = \tau^H
\end{equation}

\subsection{Spectral Analysis}
Spectra analysis is used for the estimation of possible periodic cycles in a given time series.  Atmospheric and climatic data show several periodicity such as the Quasi-Biennial Oscillation (2 - 3 years), El-Nino Southern Oscillation (5 - 7) years, Sunspot cycles (10- 15 years) and Atlantic Multi-Decadal Oscillation (30 years) \citep{Oguntunde2011}.  It is intuitive to investigate periodicity in the onset and cessation of rainfall for several reasons such as development of a model or investigation of coupling with other known cycles.   The Lomb-Scargle periodogram which has shown great promise in detecting peaks in unevenly sampled time correlated data will be used in this study \citep{clifford2005}.  The Lomb-Scargle periodogram is given as
\begin{equation}\label{lomb}
    P_N(\omega) = \frac{1}{2\sigma^2}\left\{ \frac{\left[\sum_j (x_j - \bar{x}\cos(\omega(t_j - \tau))\right]^2}{\sum_j \cos^2(\omega(t_j - \tau))} + \frac{\left[\sum_j (x_j - \bar{x}\sin(\omega(t_j - \tau))\right]^2}{\sum_j \sin^2(\omega(t_j - \tau))} \right\}
\end{equation}
$\bar{x}$ and $\sigma^2$ are the mean and variance of the time series respectively, $\tau$ is a frequency based time delay given by
\begin{equation}
    \tan(4\pi f\tau) = \sum_{n-1}^N \frac{\sin(4\pi f t_n)}{\cos(4\pi f t_n)}
\end{equation}
\subsection{Model development}
In order to capture the periodic and linear trend a model of the form

\begin{equation}\label{enq1}
   y =  \sum_{n-1}^N  A_i\cos \left(\frac{2\pi t - \theta}{T}\right) + Bt + C
\end{equation}
where N ($N = 5$) is the total number of periods considered,  t is the year, $\theta$ is the phase, T is the period determined using the Lomb-Scargle power spectral density estimate. In this work, $\theta$ was chosen as 1984 strictly for convenience.  $A_i, B, C$ are constants to be determined by nonlinear curve fitting methods.

\section{Results and Discussion}

\subsection{Linear Trend Analysis}

\begin{figure}
  \centering
  \includegraphics[width=\textwidth]{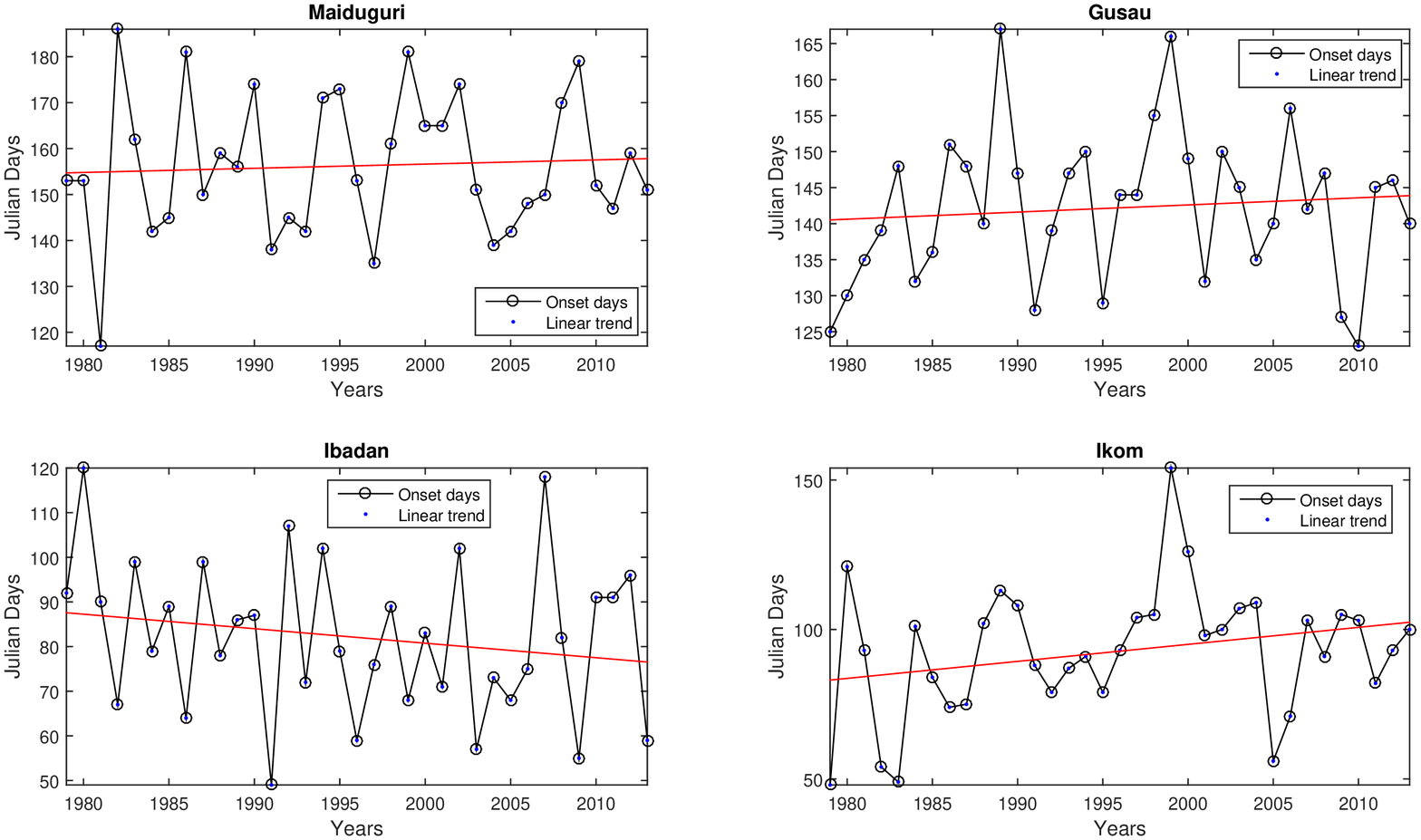}\\
  \caption{Linear trend in onset of growing season for the four locations under consideration}\label{fig2}
\end{figure}

\begin{figure}
  \centering
  \includegraphics[width=\textwidth]{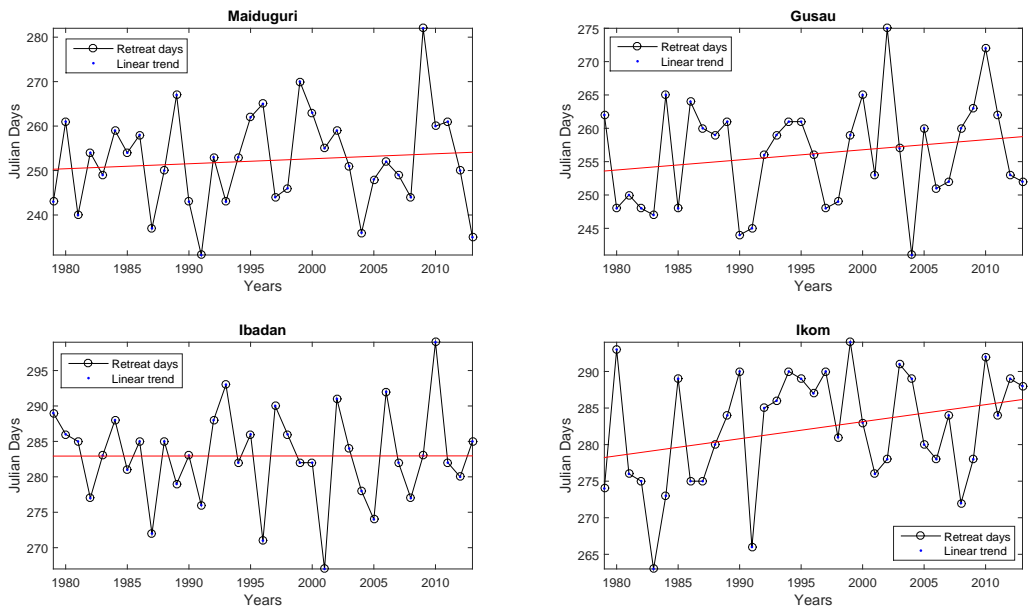}\\
  \caption{Linear trend in cessation of growing season for the four locations under consideration}\label{fig3}
\end{figure}

\begin{figure}
  \centering
  \includegraphics[width=\textwidth]{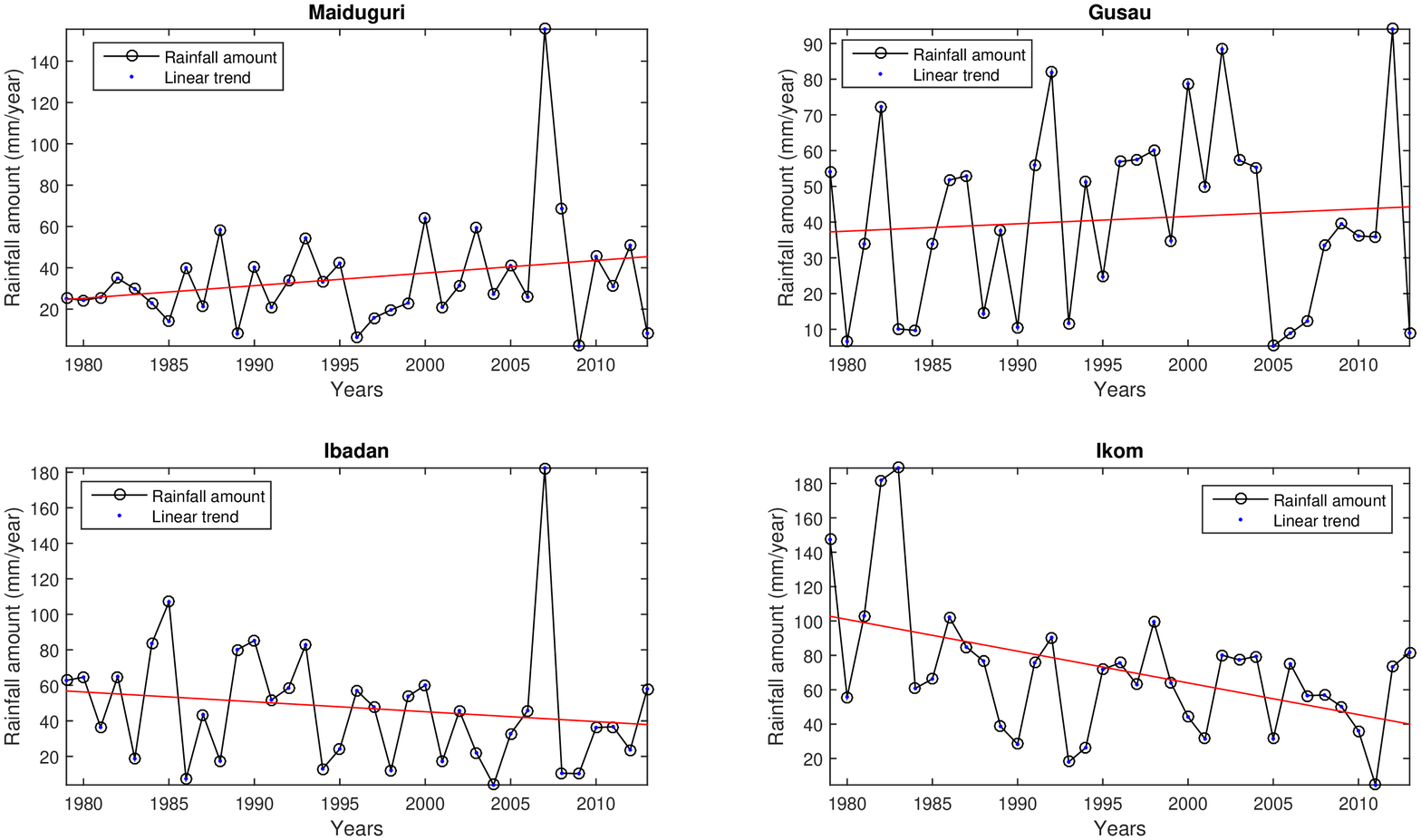}\\
  \caption{Linear trend in amount of rainfall at the onset of growing season for the four locations under consideration}\label{fig4}
\end{figure}

\begin{figure}
  \centering
  \includegraphics[width=\textwidth]{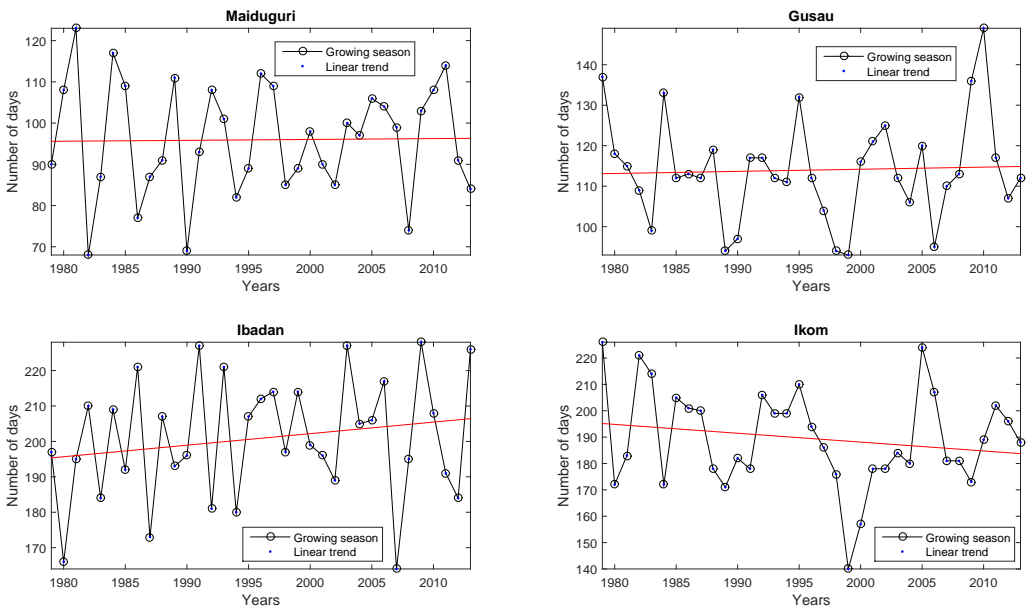}\\
  \caption{Linear trend in the length of growing season for the four locations under consideration}\label{fig5}
\end{figure}

The linear trends obtained for the onset, cessation, rainfall amount during the onset and length of the growing season are shown in Figures \ref{fig2} - \ref{fig5}  and the summary of trend statistics presented in table \ref{tab1}.  The two northern locations show onset dates in the range $120 - 190$ and $125 - 165$ for Maiduguri and Gusau respectively while Ibadan and Ikom showed onset dates in the range $50 - 120$ and $50 - 150$ respectively (Figure \ref{fig2}).  These values are very close to that reported by \citet{Oguntunde2014} using another method of computing the rainfall onset and retreat.  The early ROD in the two southern locations could be attributed to the closeness to the Atlantic Ocean and prevailing Mangrove ecological zone of the area \citep{Ogunjo2016}.  The RCD for each of the locations studied is shown in Figure \ref{fig3}. The northern stations have earlier RCD than the southern stations, similar to the observation in ROD.  Figure \ref{fig4} shows the variation of the amount of rainfall available at the onset of rainfall in each of the locations studied.  Maiduguri and Ibadan had unusually high rainfall amount at the onset of the 2007 raining season which did not reflect in the Gusau and Ikom stations.  Maiduguri and Ikom had the smallest in the highest volume of rainfall at the onset of rainfall among the four stations respectively.  The dichotomy between the northern and southern climate was also reflected in the length of the growing season (Figure \ref{fig5}).

Table \ref{tab1} shows the result obtained for linear trends in ROD, RCD, RAO and LGS.  Positive trends were observed for rainfall onset dates in all locations except Ibadan where  a negative trend is reported.  Ikom was observed to have the highest trend value of 0.569 days/year.  The slope for the two northern stations are close.  Positive slopes were obtained for rainfall cessation dates in all locations with Ikom and Ibadan having the highest and lowest values respectively.  Positive values were obtained for the rainfall amount at onset of rainfall for the northern stations while the southern stations have negative values.  The method of linear trend is susceptible to extreme values, hence, the need for more robust testing.

Table \ref{tabsen} shows result obtained for Mann-Kendall, Sen Slope and Hurst Exponent for Rainfall onset, cessation, length of the growing season and rainfall amount at the onset of rainfall.  Using the Mann-Kendall test for trend, negative trends were observed in  ROD (Ibadan), RCD (Ibadan), LGS (Gusau and Ikom) and  RAO (Ibadan and Ikom).  Using the Mann-Kendall test for trend, there were insufficient evidence to reject the null hypothesis of trend absence in all the cases (rainfall onset days, rainfall cessation days, rainfall amount at onset and length of growing season) for all location except the rainfall amount at onset for Ikom where there were sufficient evidence of trend.  The trend reported here are similar to that reported by \citet{Oguntunde2014} for the same locations except for LGS where a negative trend was obtained by \citet{Oguntunde2014} but a positive trend was observed in this study.

True slope was obtained using the method of Sen slope.  Maiduguri showed very low slopes for ROD, RCD and LGS while a value of 0.38 days/year was obtained for RAO.  Gusau has the same value for ROD, RCD, and RAO while the LGS is 0.00.  Ibadan had a small decreasing trend of -0.35 for RCD but a high decrease rate for RAO.    Ikom had the highest increase rate in ROD (0.56) and RCD (0.20) as well as the highest decrease rate in RAO (-1.27) amongst all the stations studied.  All stations showed anti-persistence in ROD, RCD and LGS except Ikom.   Similarly, persistence was observed in all stations for RAO except Ibadan which showed anti-persistence characteristics.

\subsection{Periodic Trend Analysis}\label{pta}

The Lomb-Scargle periodogram was used to estimate the first five (5) dominant frequencies for rainfall onset and cessation and the result presented in Table \ref{tabs}.  The dominant peaks for all regions in ROD and RCD (except Gusau) are in the range $3-5$.  The frequencies for Maiduguri are in the range $2 - 7$, which suggests the influence of the Stratospheric Quasi-Biennial Oscillation (2 -3 yr) and El Nino Southern Oscillation (5 - 7 yr) on the onset of rainfall in the location.  In addition to the two external drivers, the RCD shows frequencies in the range 10 - 15 which also suggest the influence of sunspot cycles (10 - 15).  Gusau show dominant frequencies for ROD and RCD in the range 3 - 7, 10 - 15, 35 yr, which also suggests the influence of the Atlantic Multi-Decadal Oscillation sea surface temperature.

The dominant frequencies in the Ibadan station is in the range 2 - 5 and 10 - 15 which can be interpreted as being driven by the Stratospheric Quasi-Biennial Oscillation and sunspot cycles respectively.  In Ikom, the dominant frequencies are in the range 3- 5, 8, 10 - 15, 20 and 35.  The predominant frequencies for ROD and RCD in the range 3 - 5 could not be accounted for using known external drivers such as Atlantic Multi-Decadal Oscillation sea surface temperature or El nino Southern Oscillation, hence, there is the need for further investigations.

\subsection{Model Development}
\begin{figure}
  \centering
  \includegraphics[width=\textwidth]{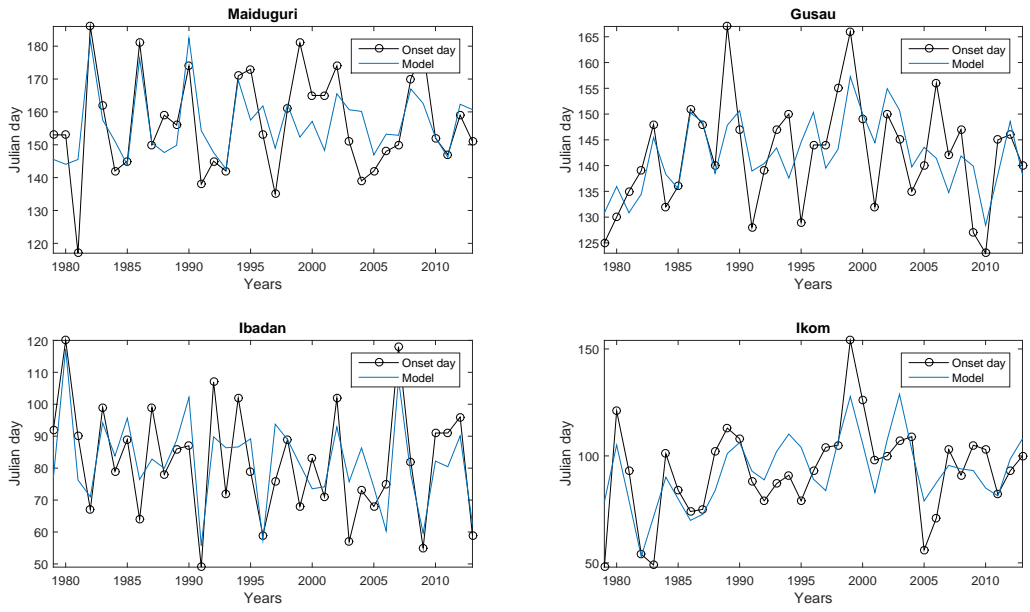}\\
  \caption{Comparison of model and data for onset of growing season for the four locations}\label{fig6}
\end{figure}

\begin{figure}
  \centering
  \includegraphics[width=\textwidth]{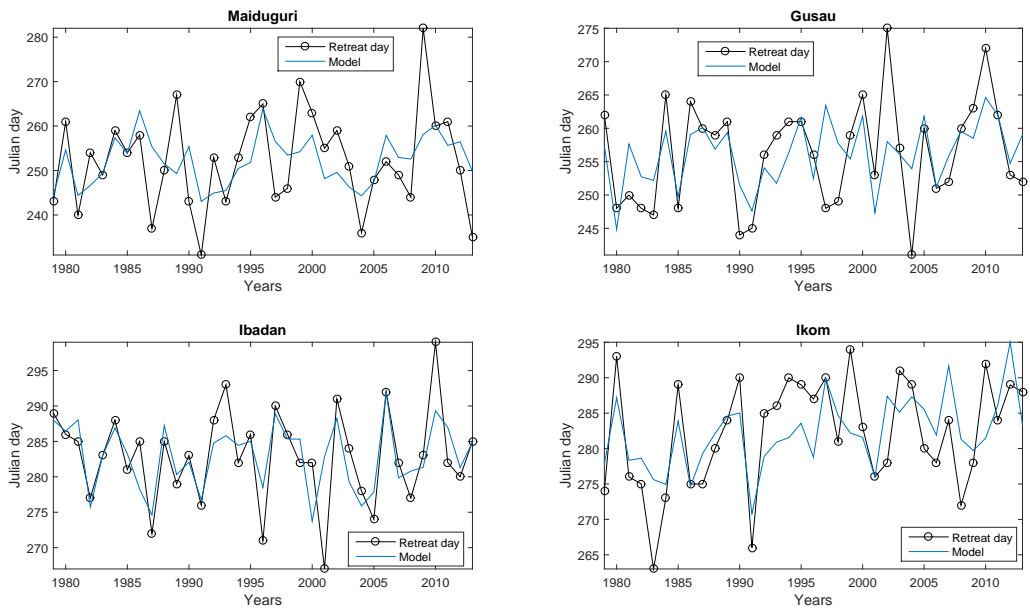}\\
  \caption{Comparison of model and data for cessation of growing season for the four locations}\label{fig7}
\end{figure}

Using the five dominant periods obtained in section \ref{pta}, a Fourier based model with linear trend accounted for as described in Equation \ref{enq1} was developed.  A comparison of the model and data is presented in Figures \ref{fig6} and \ref{fig7} for ROD and RCD respectively.  Model parameters are shown in Table \ref{tab3} and \ref{tab4} for ROD and RCD respectively.  The Mean Square Error (MSE) and Root Mean Square Error (RMSE) obtained for the models show that the models are good fit for each of the regions,  with Gusau showing the best fit.  The constant B in the developed model represents the slope of the time series.  The slopes obtained for ROD in teh developed model confirms the high value obtained in Ikom, negative value in Ibadan but not the positive value in Maiduguri using both the linear trend and Sen slope but is in agreement with the Mann-Kendall trend results.  The trend values obtained for RCD is in agreement with that obtained using linear trend with teh values sharing similar patterns. It also reflects the results obtained using Sen slope and Mann-Kendall test statistics in all locations except Ibadan.

\subsection{Implication for Agriculture}

From the result obtained using Linear trend, Mann-Kendall Test and Sen Slope, Maiduguri was found to have positive trend and slope in ROD, RCD, LCD and RAO.  The LGS and RAO were found to be increasing at the rate of 0.06 days/year and 0.38 mm/year respectively. For example, the length of the growing season would have increased by 3 days and RAO by 19mm in the next fifty years. This implies that the location will have more rainfall at the onset of rainfall and a longer growing season in future.  Hurst exponent values suggests that the trend in LGS will not always be continuous but an increasing trend will be followed by a decreasing trend while an increasing trend will always be experienced for RAO.

Results obtained from the three methods for LGS in Gusau were not consistent.  Linear regression shows positive trend, Mann-Kendall show negative trend (albeit, insignificant) while Sen Slope shows zero slope.  However, positive trends were obtained for ROD, RCD and RAO in all the three methods.  Sen slope obtained for ROD and RCD indicates that the planting season is changing at the same rate, i.e.  both ROD and RCD will come later at the same rate.  This implies that in the coming years, the planting season will be moving towards the raining season.  Farmers in the region are advised to adopt crops that thrive in much rainfall.  To avoid destruction of agricultural products during and after planting by excess rainfall, irrigation and post harvest strategies under heavy rainfall should be made and initiated.

Ibadan has been shown to have positive and negative trend for LGS and RAO respectively.  This implies that the growing season will be increasing but the amount of rainfall at the onset of rainfall will be reducing.  The ROD and RCD will be coming earlier in the year than is currently obtained.  All the changes will not be gradual, as suggested by the Hurst exponent values. The decreasing amount of rainfall could be due to the earlier ROD which pushes the onset of growing season into the late dry season at the rate of 0.35 days per year.  Furthermore, the rainfall available at the onset of the growing season will be decreasing at the rate of 0.82 mm/year which implies that much water will not be available at the beginning of the raining season.  To mitigate the effect of these expected changes on agriculture in the region, improved food crop varieties that can be grown twice a year under irrigation scheme and with little water requirements should be considered.

The LGS and RAO for Ikom both show decreasing trends.   This implies that the growing season in the location will be decreasing with significant decreasing amount of rainfall at the beginning of the raining season at the highest rate among the four locations under consideration.  As a coastal city, there will be loss of livelihood for fish-farmers due to expected drought.  The following strategies are suggested for the location to mitigate the effect of climate change on their agricultural output and livelihood: preparation for dry season farming (crop and fish farming) with large scale irrigation scheme should be initiated, adopting drought resistant crops or crops which can germinate under minimal water requirements is necessary, and enhanced crops such as short-time growing tubers and grains should be adopted in light of reducing length of growing seasons.

\section{Conclusion}
This study investigated trends in ROD, RCD,LGS and RAO over four locations in Nigeria.  The ROD and RCD were computed using the method of cumulative percentage mean rainfall.  Three different approaches were used to compute the trend values:  Hurst Exponent, linear regression and Mann-Kendall.  The slope was obtained using Sen Slope.  From the results obtained Ibadan was found to be at risk of reduced ROD, RCD and LGS while Ikom is expected to have significant increase in ROD, RCD and LGS.  Persistence was observed in RAO for all locations under consideration and Ikom showed persistence in the parameters (ROD, RCD, LGS,RAO) which indicates an increasing trend.  The implications of the results obtained on agricultural production in the locations were highlighted and plausible responses and adaptation were proposed.  The locations are likely to experience new climate in the future.  Predominant frequencies driving ROD and RCD were investigated using the Lomb-Scargle periodogram.     There is the need for research to investigate the source of the predominant frequencies (3 - 5 years) to determine the source.  Other driving frequencies identified include the Stratospheric Quasi-Biennial Oscillations, Atlantic Multi-Decadal Oscillations, El Nino and Sunspots cycle. Climate change education is necessary for the locations considered to adequately prepare the locals for the expected changes to the climate.

\clearpage
\begin{table}
  \centering
  \caption{Linear trend values.  ROD - Rainfall onset days; RCD - rainfall cessation days; LGS - length of growing season; RAO - Rainfall amount at onset of rainfall}
  \begin{tabular}{|l|c|c|c|c|}
    \hline
Location	  &   ROD (days/year)	&   RCD (days/year) &	RAO (mm/year) &	  LGS (days/year) \\ \hline
Maiduguri &   0.091	 &   0.112	& 0.61	 &  0.021\\
Gusau	  &   0.099	 &    0.152	& 0.206	 &  0.052\\
Ibadan	  &    -0.325 &	0.001	& -0.557 &	  0.326\\
Ikom	  &   0.569	  &  0.233	& -1.846 &	  -0.336\\ \hline
  \end{tabular}
  \label{tab1}
\end{table}

\begin{table}
  \centering
  \caption{Mann-Kendall test statistics (Z-test), Sen slope and Hurst Exponent using DFA for ROD - Rainfall onset days; RCD - rainfall cessation days; LGS - length of growing season; RAO - Rainfall amount at onset of rainfall. $^*$ Trend is significant at 0.05 }
  \begin{tabular}{|l|c|c|c|c|c|c|c|c|c|c|c|c|}
    \hline
   	  &  \multicolumn{4}{c|}{Sen Slope}	&  \multicolumn{4}{c|}{Z score} &  \multicolumn{4}{c|}{Hurst Exponent}	\\	 \hline
Location    & ROD	& RCD	& LGS	& RAO	& ROD	 & RCD	& LGS	& RAO & ROD	& RCD	& LGS	& RAO\\ \hline
Maiduguri	&0.00	&0.07	&0.06	&0.38	&0.10	 &0.37	 &0.16	&1.08  & 0.11 & 0.37 & 0.20 & 0.64 \\
Gusau	&0.13	&0.13	&0.00	&0.13	&0.67	 &0.94	&-0.17	 &0.58 & 0.38 & 0.28 & 0.32 & 0.70 \\
Ibadan	&-0.35	&-0.05	&0.33	&-0.82	&-1.09	 &-0.48	&0.97	 &-1.79 & 0.27 & 0.16 & 0.27 & 0.38\\
Ikom	&0.56	&0.20	&-0.25	&-1.27	&1.54	 &1.65	&-0.75	 &-2.22$^*$ & 0.68 & 0.69 & 0.54 & 0.80\\
    \hline
  \end{tabular}
  \label{tabsen}
\end{table}

\newpage

\begin{table}
  \centering
  \caption{First five predominant frequencies (yr) for rainfall onset days and rainfall cessation days arranged in order of reducing power}
  \begin{tabular}{|l|c|c|c|c|c|c|c|c|c|c|}
    \hline
          &  \multicolumn{5}{c|}{Rainfall onset} & \multicolumn{5}{c|}{Rainfall cessation}\\ \hline
Location  &$f_1$& $f_2$ & $f_3$ & $f_4$ &$f_5$ & $f_1$ & $f_2$ & $f_3$ & $f_4$ &$f_5$\\ \hline
Maiduguri &3.26	&3.89	&4.24	&6.36	&2.03	 &3.26	&5.19	 &11.67	&7.78	&2.03\\
Gusau     &3.18	&4.24	&6.09	&14.00	&35.00	 &7.78	&2.64	 &11.67	&2.33	&5.38\\
Ibadan    &4.52	&2.46	&2.22	&3.41	&3.04	 &4.38	&2.26	 &3.68	&3.41	&15.56\\
Ikom      &4.67	&10.77	&35.00	&3.78	&3.26	 &4.52	&20.00	 &8.24	&2.46	&3.68\\ \hline
  \end{tabular}
  \label{tabs}
\end{table}

\begin{table}
  \centering
  \caption{Model parameters for rainfall onset days and error statistics - Mean Squared Error (MSE) and Root Mean Squared Error (RMSE)}
  \begin{tabular}{|l|c|c|c|c|c|c|c|c|c|}
    \hline
Location    & $A_1$	& $A_2$	& $A_3$	&  $A_4$	  & $A_5$	& $B$  &	$C$  & MSE  & RMSE\\  \hline
Maiduguri	&1.934	&-7.362	&-10.031	&-4.801	 &7.449	&-0.013	 & 156.863 & 123.99 & 11.13\\
Gusau	&6.141	&-2.558	&-1.222	& -4.595	 &-4.188	&0.016 &	 142.363 & 61.01 & 7.81\\
Ibadan	&10.197	& 11.772  &	-2.637  &	4.900 &	 6.588  &	 -0.279 &	86.474 & 111.66 & 10.57\\
Ikom	&14.959	& 5.656	  &-12.072	&6.861	 &4.738	&0.418	 &84.973 & 217.05 & 14.73\\
    \hline
  \end{tabular}
  \label{tab3}
\end{table}

\begin{table}
  \centering
  \caption{Model parameters for rainfall cessation days and error statistics - Mean Squared Error (MSE) and Root Mean Squared Error (RMSE)}
  \begin{tabular}{|l|c|c|c|c|c|c|c|c|c|}
    \hline
Location    & $A_1$	& $A_2$	& $A_3$	&  $A_4$	  & $A_5$	& $B$  &	$C$ & MSE & RMSE\\ \hline
Maiduguri	&3.205	& 2.705 &	-6.053	& 0.799 &	 2.393  &	 0.028 &	251.581 & 85.17 & 9.23  \\
Gusau	    &1.851	&-4.695	&-2.831	 &-1.958	 &-1.695	& 0.116	& 254.179 & 37.98 & 6.16 \\
Ibadan	    &4.558	&-3.446	&1.854	&-0.887	 &2.076	&0.020	 &282.284 & 20.80 & 4.56 \\
Ikom	&3.221	&0.539	&2.934	&3.797	&-1.211 &	 0.262  &	 277.891 &  34.52 & 5.88 \\
    \hline
  \end{tabular}
  \label{tab4}
\end{table}

\clearpage
\bibliography{mybibfile}

\end{document}